\documentstyle[12pt]{article}
\newcommand{\tr}{\mathop{\rm tr}\nolimits}
\newcommand{\rank}{\mathop{\rm rank}\nolimits}

\begin{document}

\begin{titlepage}
\vspace*{2.5cm}

\begin{center}

{\large\bf The dark matter and the condensed description\\
 by the gravitational fields in the field theory}\\ [3mm]

\vspace{0mm}

               V.M.~Koryukin   \\
\vspace{0mm}
{\it Mari State Technical University, Lenin sq. --- 3,
 Yoshkar--Ola, 424024, Russia}   \\
 koryukin@marstu.mari.ru
\end{center}

\vspace{0mm}

\noindent

 In this work the approach for a description of the physical
 systems about whiches it is not possible to get
 basicly the total information is suggested.
 As a result the differential equations, the solutions
 of which characterize the physical system, must bieng obtained
 from the demand of the minimality of the
 generalized variance.
 As the understanding of the particles trajectory absents in the
 quantum theory we shall consider that the Riemannian
 space--time~$M_n$ is the effective one, postulating the metric
 tensor on the base of the reduced density matrix~$\rho'$ of the
 gravitational fields~$\Phi (x)$.
 It is naturally that the presence of the dark matter in the ground
 state determines the non-trivial structure of the reduced density
 matrix~$\rho'$ and the non-trivial geometric
 structure of the space--time manifold~$M_n$, correspondingly.
 What is more, define those
 mixtures~$\Phi (x)$ of the gauge fields~$B(x)$
 which have the non-zero vacuum averages as the gravitational fields
 even if the space--time~$M_n$ is not the Riemannian manifold.
 Besides we shall consider that the dimensionality~$n$ of the
 space--time is the rank of the density matrix~$\rho$ of the gauge
 fields~$B(x)$ in the ground (vacuum) state.



\end{titlepage}

 Considering the Feynman's algorithm of the path integrals we shall
 use the set of the wave functions~$\{\Psi(x)\}$ instead of the one, 
 where the point~$x\in M_n$ ($M_n$ is the space-time). Of course for
 the condensed description it is convenient to have only one
 ``theoretical'' function which we shall seek as the extremal one for
 the action (the generalized variance)
\begin{equation}
 \label{1}
 {\cal A} = \int\limits_{\Omega_n} {\cal L} d_nV ,
\end{equation}
 where~$\Omega_n\subset M_n$ and~$d_nV$ is the measure of
 the~$d_n\Omega\subset\Omega_n$. Write down the
 Lagrangian~${\cal L}$ in the form
\begin{equation}
\label{2}
 {\cal L}=\kappa\overline{X^b}(\Psi)\rho_b^a X_a(\Psi)
\end{equation}
 (what is agreed with the interpretation of the quantum field as
 the infinite set of the interacting harmonic oscilators),
 where~$a,b,c,d,e = 1,2,...,r$; $\kappa$ is a constant;
 $\rho_a^b(x)$ are the components of the density matrix~$\rho(x)$
 ($\tr\rho = 1, \rank\rho = n, \rho^+ = \rho$, the
 top index~$+$ is the Hermitian conjugation symbol); the
 bar means the Dirac conjugation symbol, which is the
 superposition of the Hermitian conjugation and the space
 inversion;
\begin{equation}
 \label{3}
 X_a(\Psi ) = T_a(\Psi) - \xi_a^i \partial_i\Psi ,
\end{equation}
 ($\partial_i$ are partial derivatives;
 $i,j,k,l,m = 1,2,...,n$; in particular, $n=4$)~\cite{k}.
 Here and further~$T_a$ are the transition
 operators, which are obtained via ``empiric'' wave functions.

 It is naturally demand that the considered Lagrangian be invariant by
 the following infinitesimal transformation
\begin{equation}
 \label{4}
 x'^i=x^i+\delta x^i \cong x^i+\delta \omega^a \xi_a^i (x) .
\end{equation}
\begin{equation}
 \label{5}
 \Psi '(x) = \Psi + \delta\Psi \cong
 \Psi + \delta\omega^a T_a(\Psi )
 \cong \Psi + \delta_o\Psi +
 \delta \omega^a \xi_a^i \partial_i \Psi ,
\end{equation}
 where
\begin{equation}
 \label{6}
 \delta_o\Psi = \delta\omega^a X_a(\Psi) .
\end{equation}
 Let
\begin{equation}
 \label{7}
 B_{\gamma}^b \overline{B^{\gamma}_a}=
 \rho_a^b B_{\gamma}^c \overline{B^{\gamma}_c} ,
\end{equation}
 then the change~$\delta_o B^a_{\beta}$ of the
 fields~$B^a_{\beta}(x)$ in the point~$x\in M_n$ must depends on
 both parameters~$\delta\omega$ and their partial
 derivatives~$\partial_i\delta\omega$. Such fields are named
 gauge ones~\cite{yan}.

 Introduce the metric in the space-time~$M_n$, using
 the reduced density matrix~$\rho '(x)$, with the components
\begin{equation}
\label{8}
 \rho_i^j = \overline{\xi_i^b}\rho_b^a\xi_a^j /
 (\overline{\xi_k^d}\rho_d^c\xi_c^k)
\end{equation}
 and considering~$M_n$ the Riemannian space-time.
 Let the fields
\begin{equation}
\label{9}
 g^{ij} = \eta^{k(i} \rho_k^{j)} (\eta_{lm} g^{lm})
\end{equation}
 (where~$\eta_{ij}$ are the covariant components of the metric
 tensor of the tangent space to~$M_n$ and $\eta^{ij}$ are
 defined as the solutions of the following equations:
 $\eta^{ij}\eta_{kj}=\delta^i_k$) are the components of the
 tensor inverse to the fundamental one of the space-time~$M_n$. As a
 result the construction of the differentiable manifold~$M_n$ can be
 connected with solving the equations for the gauge
 fields mixtures defined as
\begin{equation}
\label{10}
 \Phi_{\beta}^i = B_{\beta}^a \xi_a^i
\end{equation}
 and obtained from the minimality requirement for the total
 generalized variance
$$
 {\cal A}_t = \int\limits_{\Omega_n} {\cal L}_t d_nV =
 \int\limits_{\Omega_n} ({\cal L} + {\cal L}_1) d_nV =
$$
\begin{equation}
\label{11}
 \int\limits_{\Omega_n} [\kappa \overline{X^b} (\Psi)
 \rho_b^a(B) X_a (\Psi) + \kappa_1 \overline{Y^c{}^{\beta}_d}(B)
 \rho_1{}_{c\beta}^d{}^{a\gamma}_b(B) Y_a{}_{\gamma}^b(B)] d_n V .
\end{equation}
 where~$\kappa_1$ is a constant, $\rho_1{}_{c\beta}^d{}^{a\gamma}_b$
 are the components of the density matrix~$\rho_1(x)$,
\begin{equation}
\label{12}
 Y_a{}_{\gamma}^b(B) = T_a{}_{\gamma}^b (B)
 - \xi_a^i\partial_iB^b_{\gamma}
\end{equation}
 ($T_a{}_{\gamma}^b$ are the transition operators
 for the fields~$B^c_{\delta}$). If it give concrete expression
 to the operators~$T_a$ and~$T_a{}_{\gamma}^b$ then
 it is desirable to consider that~$X_a(\Psi)$
 and~$Y_a{}_{\gamma}^b(B)$ are the generators of the Lie local
 loop~$G_r$~\cite{k}.
 Precisely the structure of the Lie local loop will
 characterize the degree of the coherence considered by us the quantum
 system. By this the maximal degree is being reached for the Lie
 simple group and the minimal degree is being reached for the Abelian
 one. In the last case we shall have the not coherent mixture of the
 wave-functions, it's unlikely which can describe the unspreading
 wave packet that is being confirmed by the absence of the fundamental
 scalar particles, if hipothetical particles are not being taken into
 account (in experiments only the mesons, composed from the quarks,
 are being observed and which are not being considered the fundamental
 one). Note that the ``soft'' structure of the Lie local loop by
 contrast to the Lie group allow to use it by the description of the
 symmetry both the phase transition (there is the time dependence)
 and the compact objects (there is the space dependence) especially.

 We construct the differentiable manifold~$M_n$, not interpreting it
 by physically. Of course we would like to consider the manifold~$M_n$
 as the space-time~$M_4$. At the same time it is impossible to take
 into account the possibility of the phase transition of a system as a
 result of which it can expect the appearance of the coherent states.
 In consequence of this it is convenient do not fix the dimensionality
 of the manifold~$M_n$. It can consider that the macroscopic system
 reach the precisely such state by the collapse. As a result we have
 the classical analog of the coherent state of the quantum system.
 Besides there is the enough developed apparatus --- the dimensional
 regularization using the spaces with the changing dimensionality and
 representing if only on the microscopic level.

 From the recent experimental data (see for example~\cite{TZH}) it
 is followed that only 5\% of the all Universe matter has the baryon
 natture, 33\% is the dark matter and 62\% exists in the vacuumly similar
 state ($p = - \rho$, where $p$ is the pressure, $\rho$ is the energy
 density) which is connected with the $\Lambda$-term. In cousequence of
 this it is expedient to divide the Universe matter into the rapid and
 slow subsystems considering that all known particles (ignoring
 neutrinos) fall into the rapid subsystem and using the
 fields~$\Phi(x)$ with the non-zero vacuum averages for the condensed
 description of the slow subsystem. Thus the elemental particles can
 be cosidered as the coherent structures in the open systems
 characterized if only the quasigroup symmetries~\cite{k}.

 Naturally, that the assumption about fields are filling the
 Universe and determining the geometric structure of
 the space--time manifold, allows one to introduce the connection of the
 fundamental tensor of this manifold with such a statistical
 characteristic as the entropy defining it in a standard manner via the
 reduced density matrix~$\rho' (x)$ in the form
\begin{equation}
\label{13}
  S = - \tr (\rho' \ln \rho' ) .                                     
\end{equation}
 As a result, the transition from the singular state of the Universe to
 the modern one can be connected with increasing of the entropy~$S$
 defined here. As~$1<n<r$, we can consider that the gauge
 fields~$B_{\beta}^a(x)$ became the owners of the nonzero vacuum
 averages breaking their symmetry in consequence of the Universe
 evolution. The presence of the nonzero vacuum averages lead to that
 the fields equations
\begin{equation}
\label{14}
 \Phi_{\alpha}^j~\left(\frac{{\cal L}_t}{\eta}~\frac{\partial
 \eta}{\partial B_{\alpha}^b} + \frac{\partial {\cal L}_t}
 {\partial B_{\alpha}^b} -
 \nabla_i\left(\frac{\partial {\cal L}_t}
 {\partial\nabla_i B_{\alpha}^b}\right)\right)
 B_{\beta}^b \Phi_k^{\beta} = 0
\end{equation}
 ($\nabla_i$ are covariant derivatives, $\eta (x)$ is the
 base density of the space-time~$M_n$)
 are not converted to zero identically by freezing of the
 excitations being the quanta of the fields~$\Phi^i_{\alpha}(x)$
 but they convert to the Einstein equations or to their
 generalizations~\cite{k}. In consequence of this we consider
 that General Relativity are only the condensed description of the
 gravitational phenomena.

 There is the prevailing opinion in the theoretical physics, that
 gravitational interactions can be neglected on the microscopic
 level. In consequence of this gravitational fields are not being
 enlisted for the description of the interaction of elemental
 particles. We shall
 consider that even in the case if it can neglect the gradients of the
 gravitational fields on account of their possible homogeneity but
 this fields it is not allowed to neglect in consequence of 
 quantum effects what is showed in the presence of the spins and
 the masses of elemental particles.


\begin{thebibliography}{999} \itemsep=-5pt

\bibitem{k}
 {\normalsize V.M.~Koryukin, Gravitation and
 Cosmology {\bf 5}, No. 4 (20), 321 (1999).}
\bibitem{yan}
 {\normalsize C.N.~Yang, R.L.~Mills, Phys. Rev. {\bf 96},
              191 (1954).}
\bibitem{TZH}
 {\normalsize M.~Tegmark, M.~Zaldariaga, A.S.~Hamilton,
  Phys. Rev. D {\bf 63}, 043007 (2001).}

\end{thebibliography}
\end{document}